\documentstyle[12pt, titlepage, cite, epsf, rotate]{article}
\def\_{\rule{.3em}{.15ex}}
\newcommand{\be}{\begin{equation}}
\newcommand{\ee}{\end{equation}}
\newcommand{\bea}{\begin{eqnarray}}
\newcommand{\eea}{\end{eqnarray}}
\newcommand{\f}{\frac}
\newcommand{\ra}{\rightarrow}

\newcommand{\al}{\alpha_s}

\newcommand{\bsg}{$b \ra s \gamma$ }
\newcommand{\Bsg}{$B \ra X_s \gamma$ }
\newcommand{\MS}{$\overline{MS}$ }
\def\slash#1{\setbox0=\hbox{$#1$}#1\hskip-\wd0\dimen0=5pt\advance
       \dimen0 by-\ht0\advance\dimen0 by\dp0\lower0.5\dimen0\hbox
         to\wd0{\hss\sl/\/\hss}}
\begin{document}
\begin{titlepage}

 \begin{flushright}
  {\bf TUM-T31-79/94}\\
    hep-ph 9409454\\
   September 1994
 \end{flushright}

\begin{center}
\vspace*{0.6in}
\renewcommand{\thefootnote}{\fnsymbol{footnote}}
\setcounter{footnote}{0}
\setlength {\baselineskip}{0.3in}
  {\Large\sf Two-loop mixing of dimension-five flavor-changing
operators}\footnote{Supported by the German Bundesministerium f\"ur
Forschung und Technologie under contract 06 TM 732, by the CEC Science
project SC1-CT91-0729 and by the Polish Committee for Scientific
Research.}
\vspace{1.2in} \\
\setlength {\baselineskip}{0.2in}

{\large\sf Miko\l aj Misiak}\footnote{On leave of absence from the Institute of
Theoretical Physics, Warsaw University.}
{\large and }{\large\sf Manfred M\"unz}
\vspace{0.3in}

{\sl Physik-Department, Technische Universit\"at M\"unchen\\
D-85747 Garching, Germany\\}
\vspace{1.5in}

{\bf Abstract \\}
  \end{center}
\noindent
We calculate the two-loop QCD mixing of the dimension-five
flavor-changing operators that arise in the Standard Model after
integrating out the $W$, $Z$ and Higgs bosons and the top-quark,
i.e.~the mixing among gluonic and photonic magnetic moment
operators. Our calculation completes the two-loop anomalous dimension
matrix of operators that govern low energy flavor-changing
processes. It is an important ingredient of the next-to-leading
calculation of the \Bsg decay rate.
\vspace{0.5in}

\end{titlepage}

\renewcommand{\thefootnote}{\arabic{footnote}}
\setcounter{footnote}{0}
\setlength {\baselineskip}{0.3in}

\section{Introduction}

	Flavor changing processes at energies much below the
electroweak scale are most conveniently described with help of the
effective hamiltonian \cite{topstory}
\be
H_{eff} \sim \sum_i C_i(\mu) O_i(\mu)
\ee
where $O_i$ are dimension-five and -six operators built out of light
($u, d, s, c$ and $b$) quarks, leptons, photons and gluons, and
$C_i(\mu)$ are their Wilson coefficients.  The operators $O_i$ that
remain after applying the equations of motion (EOM) \cite{Grin} are
dimension-six four-fermion operators and the dimension-five magnetic
moment operators:
\bea
\label{ogamma}
O_{\gamma} & = & e \bar{q} \sigma^{\mu \nu} F_{\mu \nu} q' \\
\label{og}
O_g & = & g \bar{q} \sigma^{\mu \nu} G^a_{\mu \nu} T^a q'.
\eea
Here $F_{\mu \nu}$ and $G^a_{\mu \nu}$ denote the photonic and gluonic
field strength tensors, respectively, and $T^a$ are the color group
generators. The quarks $q$ and $q'$ have different flavor and usually
have specific chiralities but, as explained below, this is irrelevant
in our calculation. The QED and QCD coupling constants are denoted by
$e$ and $g$, respectively.

	The factorization of short- and long-distance QCD effects is
achieved \cite{topstory} by evolving the coefficients $C_i(\mu)$ from
the electroweak scale down to the low energy scale of the process
under consideration. The anomalous dimension matrix that governs this
evolution is known up to two loops for the QCD mixing of the
phenomenologically important 4-quark operators into themselves
\cite{66}, quark-lepton operators \cite{ql} and the magnetic moment
operators \cite{Grin,Pisa,bsgmix}. The only remaining two-loop QCD
mixing among all these operators is the two-loop mixing of the
magnetic moment operators (\ref{ogamma}) and (\ref{og}) which we
calculate in the present paper.

	Our results can have many phenomenological applications, as
they affect all possible flavor-changing processes involving two
quarks and a photon or a gluon. The two-loop mixing we calculate can
also be directly applied to flavor-conserving magnetic moment
operators (see end of section \ref{two-loop}). However, the most
important application of our results is their use in the calculation
of the complete next-to-leading logarithmic QCD corrections to the
\Bsg decay rate. The critical phenomenological importance of
calculating the next-to-leading QCD corrections to the \Bsg rate has
been stressed in the previous paper \cite{bh}. We will not repeat this
discussion here. Instead, we will put more emphasis on the details of
the present calculation.

	The paper is organized as follows: In the next section, we
recalculate the one-loop mixing of $O_{\gamma}$ and $O_g$. We recover
also the nonphysical one-loop counterterms that will be needed for
cancellation of subdivergences in the further two-loop computation.
Section 3 presents certain details of the two-loop calculation. Our
final results are given in the end of this section. In the last
section, we rewrite our results in the notation common for the
analyses of the \bsg decay and show the size of the next-to-leading
contributions we have calculated.

\section{The one-loop calculation}

	Our starting point is the lagrangian density
\be \label{L}
{\cal L} = {\cal L}_{QCD \times QED} + X \left( C_{\gamma} O_{\gamma} +
C_g O_g + \sum_k \tilde{C}_k \tilde{O}_k \right).
\ee
Here ${\cal L}_{QCD \times QED}$ is the full $QCD \times QED$
lagrangian for f massless\footnote{Introducing fermion masses would
not affect the mixing among $O_{\gamma}$ and $O_g$ calculated here.}
flavors, including the gauge fixing and ghost terms. X denotes some
normalization constant that does not get QCD-renormalized. The
operators $\tilde{O_k}$ are additional dimension-five operators
required as counterterms for off-shell Green's functions with
insertions of $O_{\gamma}$ and $O_g$. According to the standard
textbook by Collins \cite{Collins}, these operators either vanish by
EOM or are BRS-exact (i.e.~are BRS variations of some other
operators), and are therefore nonphysical. In the following, we will
assume nothing about these operators but just recover them from the
explicit one-loop calculation.

\begin{figure}[htb]
\centerline{
\hspace{-1in}
\rotate[r]{
\epsfysize = 3in
\epsffile{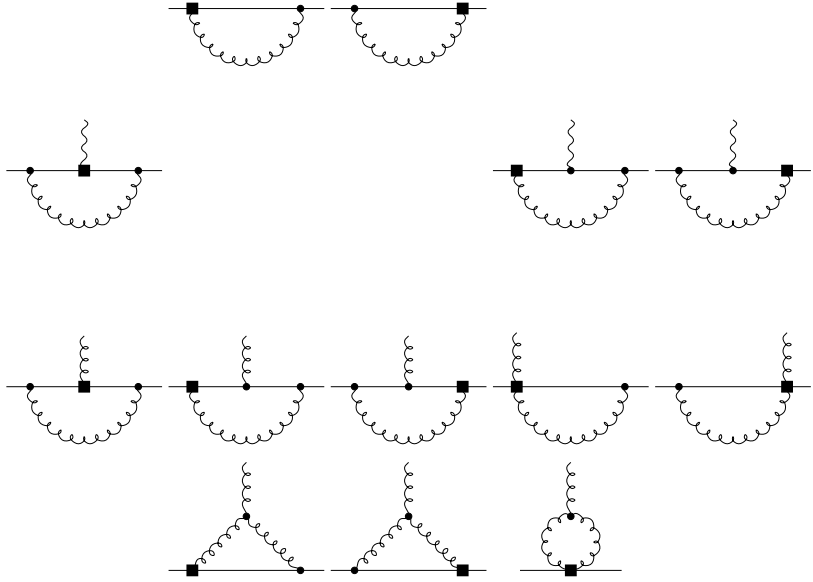}
}}
\vspace{-1in}
\caption{One-loop diagrams for the $q' \ra q$, $q' \ra q \gamma$ and
$q' \ra q Q$ transitions. The squares denote insertions of
$O_{\gamma}$ or $O_g$.}
\label{1loop.fig}
\end{figure}

	We have performed our calculation off-shell in the background
field \cite{Abbott} version of the Feynman--'t~Hooft gauge. We have
calculated UV-divergent parts of one-particle-irreducible (1PI)
one-loop diagrams with single $O_{\gamma}$ or $O_g$ insertions. All
such diagrams with nonnegative degree of divergence correspond to the
following transitions: $q' \ra q$, $q' \ra q \gamma$, $q' \ra q Q$,
$q' \ra q A$, $q' \ra q \gamma \gamma$, $q' \ra q \gamma Q$, $q'
\ra q \gamma A$, $q' \ra q Q Q$, $q' \ra q Q A$ and $q' \ra q A A$,
where Q and A denote the quantum and the background gluon fields,
respectively. The diagrams corresponding to the first three
transitions are shown in fig.~\ref{1loop.fig}.

	Since QCD and QED interactions preserve chirality, it is
obvious that the mixing we calculate is independent of whether the
operators $O_{\gamma}$ and $O_g$ are taken as they stand in
eqs.~(\ref{ogamma}) and (\ref{og}) or specific (opposite) chiralities
for the quarks $q$ and $q'$ are chosen, as it happens in the
phenomenologically important case of the \bsg decay. In the following,
we will take these operators as they stand in eqs.~(\ref{ogamma}) and
(\ref{og}).

	We find that divergences in all the possible one-loop 1PI
Green's functions with single $O_{\gamma}$ or $O_g$ insertions can be
removed by counterterms proportional to $O_{\gamma}$ and $O_g$
themselves, and to the following two operators:
\bea \label{odd}
\tilde{O}_{DD} & = & \bar{q} \slash D  \slash D q' \\
\tilde{O}_{QD} & = & i g \bar{q} \Big( -\stackrel{\leftarrow}{\slash D}
\slash Q + \slash Q \slash D \Big) q',
\eea
where $D_{\mu}$ is the usual $QCD \times QED$ covariant derivative.

	Assembling the four operators into the vector $( O_{\gamma}, O_g,
\tilde{O}_{DD}, \tilde{O}_{QD})^T$, the corresponding one-loop
anomalous dimension matrix reads
\be \label{1loop.anom}
\hat{\gamma} = \f{g^2}{16 \pi^2} \left( \begin{array}{cc|cc}
2 C_f & 0 & 0 & 0\\
8 C_f Q_q & 10 C_f - 4 N & -12 C_f & \f{3}{2} N\\
\cline{1-4}
0 & 0 & ? & ?\\
0 & 0 & ? & ? \\
\end{array} \right) + {\cal O}(g^4)
\ee
where $N$ denotes the number of colors, $C_f=(N^2-1)/2N$ and $Q_q$ is
the electric charge of the $q$-quark (which must be identical to the
one of $q'$). The mixing of $O_{\gamma}$ and $O_g$ among themselves,
as well as the mixing of $O_g$ into $\tilde{O}_{DD}$ are in agreement
with what has been found previously \cite{SVZ,Pisa}. Note that in
refs.~\cite{SVZ,Pisa} and in most other papers, the quark mass is
introduced in the normalization of $O_{\gamma}$ and $O_g$. So
$\hat{\gamma}_{\rm here} = \hat{\gamma}_{\rm there} + \gamma_m
\hat{1}$, where $\gamma_m = -\f{g^2}{16 \pi^2}6 C_f + {\cal
O}(g^4)$.

	The mixing of $O_g$ into $\tilde{O}_{QD}$ has not been
considered so far because it arises from one-loop diagrams with
external quantum gluon legs. We need to consider such diagrams as they
are divergent subdiagrams in the further two-loop calculation.

	One may ask why only the very operators $\tilde{O}_{DD}$ and
$\tilde{O}_{QD}$ arise as nonphysical counter\/terms. We know
\cite{Collins} that all such counterterms either have to vanish by the
EOM or be BRS-exact. Both $\tilde{O}_{DD}$ and $\tilde{O}_{QD}$ vanish
by the EOM for the quarks. In our case, there is not ``enough
dimension'' for the EOM for the gauge bosons (such EOM has dimension 3
and the two quark fields have also dimension 3). So any EOM-vanishing
operator must be proportional to the EOM for a quark. Two quark fields
and a covariant derivative together have dimension 4. The remaining
one unit of dimension can be provided either by another covariant
derivative (then the only possibility is $\tilde{O}_{DD}$) or by the
quantum gluon field (as in $\tilde{O}_{QD}$). The background gluon
field cannot show up because our operators must be invariant under the
gauge transformation given in eqs.~(2.14) and (2.15) of
ref.~\cite{Abbott}:\footnote{In our conventions, the sign of $g$ is
opposite to that in ref.~\cite{Abbott}.}
\bea
\delta A^a_{\mu} & = & -f^{abc} \omega^b A^c_{_\mu} - \f{1}{g}
\partial_{\mu} \omega^a \\
\delta Q^a_{\mu} & = & -f^{abc} \omega^b Q^c_{_\mu}
\eea
where $\omega^a$ denotes the gauge transformation parameter. The two
above transformations combine to the usual gauge transformation for
the full gluon field $A^a_{\mu} + Q^a_{\mu}$.

	The final observation is that our initial operators
$O_{\gamma}$ and $O_g$ are odd under charge conjugation combined with
exchange of $q$ and $q'$. This is why the two terms in
$\tilde{O}_{QD}$ come in a combination odd under this symmetry.

\begin{figure}[htb]
\centerline{
\hspace{-1.5in}
\rotate[r]{
\epsfysize = 0.5in
\epsffile{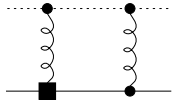}
}}
\vspace{-0.5in}
\caption{Convergent diagram for the $q' \ra q \bar{\eta} \eta$ transition.}
\label{ghost.fig}
\end{figure}

	BRS-exact operators cannot arise as counterterms to
$O_{\gamma}$ and $O_g$. The arguments for this are the following: The
BRS variation raises both dimension and ghost number by one unit. So
any BRS-exact operator that could mix with $O_{\gamma}$ or $O_g$
should be a BRS variation of a dimension-four operator with ghost number
$-1$. The only possibility for the latter operator is $\bar{\eta}^a
\bar{q} T^a q'$. The BRS variation of such an operator contains a term
with two fermions and two ghosts. On the other hand, such a term
cannot arise as a counterterm for the magnetic moment operator
because the corresponding diagram in fig.~\ref{ghost.fig} is
convergent.\footnote{At first glance, this diagram has zero
degree of divergence (remember that the operator vertex has dimension
1). But one power of momentum in the power counting is only the
momentum of the outgoing external ghost --- just because of the
structure of the gluon-ghost vertex. So the effective degree of
divergence is $-1$. This argument applies to all orders.}

	As follows from the above considerations, one can show that
$O_{\gamma}$, $O_g$, $\tilde{O}_{DD}$ and $\tilde{O}_{QD}$ are the
only possible counterterms for $O_{\gamma}$ and $O_g$ without
calculating a single Feynman diagram. Then the matrix in
eq.~(\ref{1loop.anom}) can be found just from the diagrams depicted in
fig.~\ref{1loop.fig} with only quantum gluon on external lines. The
calculation is then identical to the one in the usual Feynman--'t~Hooft
gauge (without background field).

	Since our calculation was fully computerized, it did not
require much effort to check that indeed all the divergences in the
diagrams with single $O_{\gamma}$ and $O_g$ insertions (also diagrams
with two external gauge bosons) can be renormalized by the same
counterterms proportional to $O_{\gamma}$, $O_g$, $\tilde{O}_{DD}$ and
$\tilde{O}_{QD}$. This was one of the cross-checks in our calculation.

	The last remark in order is that the operator $\tilde{O}_{QD}$
would arise also in the non-background-field calculation. Then $Q$ in
$\tilde{O}_{QD}$ would stand for the full gluon field. Consequently,
this operator would not be gauge-invariant (not even BRS-invariant).
However, this is not a problem because EOM-vanishing operators are
nonphysical anyway, i.e.~their physical matrix elements vanish. They
do not mix into physical operators, either \cite{Collins}. This is why
the unknown entries in the lower right corner of the matrix in
eq.~(\ref{1loop.anom}) are completely irrelevant.

	One may then wonder why we have decided to recover of the
mixing of $O_g$ into $\tilde{O}_{DD}$ and $\tilde{O}_{QD}$ at all. The
reason for this is the method we have applied for calculating the
two-loop diagrams in the next section.

\section{The two-loop calculation}
\label{two-loop}
	At the two-loop level we need to calculate the $q' \ra q
\gamma$ and $q' \ra q A$ Green's functions with single $O_{\gamma}$
or $O_g$ insertions. Their divergent parts give the two-loop mixing
among these operators. As a by-product, we obtain the two-loop mixing
of $O_g$ into $\tilde{O}_{DD}$. The latter mixing must come out the
same both from photonic and gluonic Green's functions. This is a
nontrivial cross-check of the calculation.

	The main problem in the two-loop calculation is the large
number of diagrams to be considered. For instance, there are 167
two-loop diagrams for the self-mixing of $O_g$. Therefore, a simple
systematic algorithm is necessary to make the calculation fully
computerized. With such an algorithm it is not really important
whether one has to calculate 100 or 1000 two-loop diagrams --- it is
only a matter of computer time which in our particular case was around
two days (on a SUN ELC workstation).

\begin{figure}[htb]
\rotate[r]{
\epsfysize = 2in
\epsffile{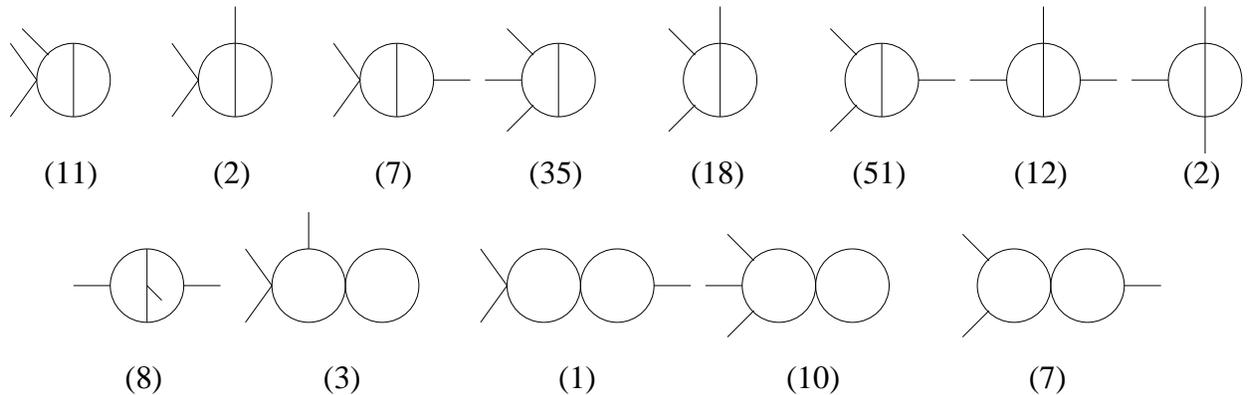}
}
\vspace{-4.5in}
\caption{Topologies of the Feynman diagrams for the two-loop
self-mixing of $O_g$. The numbers in brackets are the numbers of
diagrams corresponding to each topology.}
\label{topol.fig}
\end{figure}

	The input to our programs were the Feynman rules following
from the Lagrangian in eq.~(\ref{L}). The program {\it FeynArts}
\cite{FeynArts} was used for constructing the relevant topologies (see
fig.~\ref{topol.fig}), drawing the diagrams and creating the analytic
expressions for the corresponding amplitudes.

	Next, a single algorithm (around ten pages of {\it
Mathematica} \cite{Mathematica} code\footnote{We had two programs
created independently by each of us.}) was applied to each of the
diagrams. The potentially UV-divergent expressions were expanded to
first order in external momenta and then the tensor integrals were
reduced to scalar ones. So all the integrals we needed to calculate
were scalar two-loop integrals independent of external momenta.

	There were no physical masses in our calculation. But before
the expansion in the external momenta was made, an auxiliary mass was
introduced in all the propagator denominators. Otherwise expanding in
external momenta would create spurious infrared divergences that would
be indistinguishable from the UV-divergences within the framework of
dimensional regularization. \newpage \noindent These infrared
divergences would not cancel in the sum of all the
diagrams.\footnote{The simplest way to convince oneself about this is
to consider an example with only one relevant diagram, e.g.~one-loop
wave-function renormalization in the scalar massless $\lambda \phi^3$
theory.}

	The auxiliary mass was the same for all the propagator
denominators. So all the two-loop scalar integrals we needed to
calculate were of the form
\be \label{abc}
I_{abc} = \int \f{d^D q}{(2 \pi)^D} \f{d^D r}{(2 \pi)^D}
          \f{1}{[q^2 - m^2]^a [r^2 - m^2]^b [(q+r)^2 - m^2]^c}
\ee
where $a$, $b$ and $c$ are integers and D is the dimensionality of
spacetime. A change of variables $(q \ra q+r, r \ra -r)$ shows that
$I_{abc}$ is symmetric under permutations of $a$, $b$ and $c$. If any
of these integers is nonpositive, the integral is easily reduced to a
product of one-loop integrals. For $a$, $b$ and $c$ all positive one
has
\bea \label{111}
I_{111} & = & -\f{3 m^2}{(4\pi)^4} \left[ \f{2}{\epsilon^2} + \f{1}{\epsilon}
\left( 3 - 2 \gamma_E - 2 \ln \f{m^2}{4 \pi} \right) \right] + {\rm
convergent\;\;terms} \\
\label{211}
I_{211} & = & -\f{1}{(4\pi)^4} \left[ \f{2}{\epsilon^2} + \f{1}{\epsilon}
\left( 1 - 2 \gamma_E - 2 \ln \f{m^2}{4 \pi} \right) \right] + {\rm
convergent\;\;terms} \\
\label{n11}
I_{n11} & = & \f{(-1)^{n+1}}{(4\pi)^4 \epsilon} \f{2}{(n-1) (n-2)
m^{2(n-2)}} + {\rm convergent\;\;terms,\;\;\;\;\; for} \;\;\;\;\; n \geq 3.
\eea
where $\epsilon = 4-D$. When $a$, $b$ and $c$ are positive and two of
them are greater than 1, then $I_{abc}$ is convergent. This fact and
the above three equations are derived in the appendix.

	Eqs.~(\ref{111})--(\ref{n11}) together with similar
expressions for one-loop scalar integrals reduce the calculation of
the divergent part of any integral $I_{abc}$ in eq.~(\ref{abc}) to
simple algebra. The further steps of the calculation are of course
purely algebraic, too: The results of the scalar integrals were
combined back into tensor integrals and then contracted with the Dirac
matrices, partly using {\sc Tracer}\cite{JamLaut}. Performing the
Dirac algebra was straightforward in our case, as we had only a single
open fermion line and no $\gamma_5$ in our calculation. So no
``evanescent operators'' \cite{Dugan} had to be introduced.

	Before presenting our final results, we should give arguments
why introducing an auxiliary mass in all the propagator denominators
and then expanding in external momenta leads us to the same results
for the divergent part of the sum of all the
diagrams\footnote{i.e.~the sum of the two-loop diagrams and the
one-loop counterterm diagrams which remove subdivergences} as we
would get without introducing this auxiliary mass and without
expanding.

	The key point is the well-known fact that once all the
subdivergences are removed in a mass independent renormalization
scheme, the pole part part of a dimensionally regularized Feynman
diagram is polynomial in masses and external momenta. This fact can be
shown on the basis of given Feynman rules alone, and no
field-theoretical arguments have to be used. The proof
\cite{Collins} is based on Weinberg's theorem \cite{Weinberg} and
on the property, that differentiation with respect to external momenta
decreases the degree of divergence of a Feynman diagram with no
subdivergences. Adding auxiliary masses in propagator denominators
does not affect this proof at all. So, after introducing these
masses, the pole part of a Feynman diagram with subtracted
subdivergences still will be a polynomial in external momenta and all
the masses (including the auxiliary ones). Therefore, setting the
auxiliary masses to zero in the end, we recover the result for the
pole part of the diagram with no auxiliary masses. However, the latter
is true only for the {\em pole} part and only {\em after} subtraction
of subdivergences.

	The one-loop diagrams with counterterm vertices that cancel
the subdivergences in two-loop diagrams have to be calculated with
exactly the same Feynman rules as the two-loop diagrams. The same
refers to the calculation of the one-loop renormalization constants
themselves. So we also need to use massive propagators e.g.~in all the
diagrams for the one-loop renormalization of the two-point Green's
function for the gluon. This necessarily leads to introduction of a
``gluon mass'' that gets renormalized at the one-loop level. It is
known \cite{Zinn-Justin} that a theory with a bare mass for a vector
boson has no physical sense.\footnote{Unless this boson couples only
to conserved currents.} But so long as the would-be gauge-fixing term
is present in the Lagrangian, it is renormalizable by power counting
\cite{Itzykson-Zuber}, i.e.~the Green's functions (even if physically
meaningless) can be rendered finite with help of local
counterterms. Moreover, in a mass-independent renormalization scheme
(such as the \MS scheme we use), these counterterms will be exactly the
same as in the massless case (except for the renormalization of the
vector boson mass itself). So the Feynman rules for the physically
meaningless theory with a massive gluon can be used for recovering
renormalization constants in true QCD.

	Similar arguments can be used to justify that we could
introduce an auxiliary mass in all fermion propagator denominators
without introducing it also in the numerators. Needless to say, this
also affects the calculation of the ``gluon mass'' counterterm.

	Introducing auxiliary masses does not affect the results for
the one-loop mixing summarized in eq.~(\ref{1loop.anom}). Neither it
affects the usual QCD renormalization constants (except for
renormalization of mass terms). However, it affects the role the
EOM-vanishing operators $\tilde{O}_{DD}$ and $\tilde{O}_{QD}$ play in
our calculation. With the modified Feynman rules, the usual arguments
which show that on-shell matrix elements of these operators vanish no
longer hold. These operators remain nonphysical in the end, when we
set the auxiliary masses to zero. But they are important at the
intermediate stages of our calculation, i.e.~the one-loop counterterm
diagrams with counterterms proportional to $\tilde{O}_{DD}$ and
$\tilde{O}_{QD}$ do affect the mixing among the physical operators
$O_{\gamma}$ and $O_g$.

	A remark concerning expansion in external momenta is now in
order. As the two-loop diagrams for $q' \ra q \gamma$ and $q' \ra q A$
transitions have degree of divergence equal to $+1$, the pole parts of
our diagrams are linear in external momenta. This is why we expand in
external momenta only to the first order before performing the loop
integrations. The expansion in external momenta can be viewed as an
exact splitting of propagators into parts that are polynomial in
external momenta and parts that contribute to integrands with lower
degree of divergence, as in the following example:
\be
\f{1}{(q+p)^2-m^2} = \f{1}{q^2-m^2} + \f{-2qp-p^2}{q^2-m^2} \f{1}{(q+p)^2-m^2}
\ee
where $p$ denotes the external momentum and $q$ is the loop momentum.
Performing such an operation appropriately many times (in a diagram
with subtracted subdivergences) one can split the integral into a
convergent part and a part that is a polynomial in external momenta.
This shows that for the pole part of a diagram with subtracted
subdivergences, the expansion in external momenta is allowed before
the loop integration is made.

	Our final results were subject to several cross-checks. First,
the double pole part of the sum of all the two-loop diagrams had to be
twice smaller and have the opposite sign than the double-pole part of
the sum of all the one-loop diagrams with counterterm vertices. Only
then the sum of all the diagrams together can be local
(i.e.~polynomial in external momenta). Second, the pole parts needed
to have the structure of the gauge-invariant operators $O_{\gamma}$,
$O_g$ and $\tilde{O}_{DD}$ (the operator $\tilde{O}_{QD}$ could not
appear as we have not calculated two-loop diagrams with an external
quantum gluon). This led to relations between coefficients at
different Lorentz structures (usually two relations for four
independent structures). Next, the mixing of $O_g$ into
$\tilde{O}_{DD}$ had to come out the same from both $q' \ra q \gamma$
and $q' \ra q A$ diagrams. This is also a manifestation of gauge
invariance. It is interesting to note that the latter requirement was
fulfilled only after proper inclusion of the ``gluon mass'' one-loop
counterterms. Finally, the two-loop self mixing of $O_{\gamma}$ which
is given by a relatively small number of diagrams has been calculated
also ``by hand'' without introducing auxiliary masses and expanding in
the external momenta.

	The final result of our two-loop calculation is the explicit
expression for the ${\cal O}(g^4)$ part of the matrix $\hat{\gamma}$
in eq.~(\ref{1loop.anom}). The anomalous dimension matrix now reads
\be \label{2loop.anom}
\hat{\gamma} =
\f{g^2}{16 \pi^2}
\left( \begin{array}{cc|cc}
2 C_f & 0 & 0 & 0\\
8 C_f Q_q & 10 C_f - 4 N & - 12 C_f & \f{3}{2} N\\
\cline{1-4}
0 & 0 & ? & ?\\
0 & 0 & ? & ? \\
\end{array} \right) +
\f{g^4}{(16 \pi^2)^2}
\left( \begin{array}{cc|cc}
a & 0 & 0 & 0\\
b & c & d & ? \\
\cline{1-4}
0 & 0 & ? & ?\\
0 & 0 & ? & ? \\
\end{array} \right) +
{\cal O}(g^6),
\ee
where
\bea
\label{a}
a & = & C_f \left(\f{257}{9} N - 19 C_f - \f{26}{9} f\right), \\
\label{b}
b & = & C_f \left(\f{404}{9} N - 32 C_f - \f{56}{9} f\right) Q_q,\\
\label{c}
c & = & -\f{299}{9} -\f{51}{4 N^2} + \f{247}{36} N^2 +
\f{41}{9} \f{f}{N} + \f{2}{9} f N
\eea
\vspace*{-4mm}and\vspace*{-3mm}
\bea
\label{d}
d & = & C_f \left(-\f{643}{6} N + 60 C_f +
\f{44}{3} f\right).\phantom{Q}
\eea

\begin{figure}[htb]
\centerline{
\rotate[r]{
\epsfysize = 1in
\epsffile{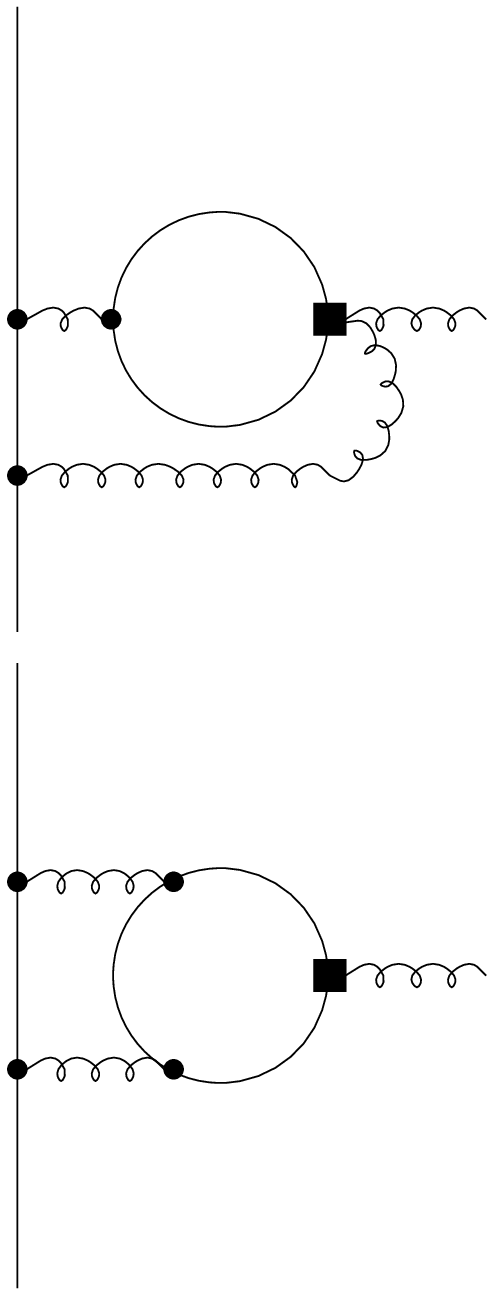}
}}
\vspace{-1.1in}
\caption{Examples of additional diagrams for flavor-conserving operators.}
\label{fig4}
\end{figure}

	Let us now briefly come to the case of two identical quark
flavors, $q\equiv q'$. Then we have to take into account several additional
diagrams of the type depicted in fig.~\ref{fig4}. All these diagrams
contain a trace over an odd number of Dirac matrices (remember we are
working with massless quarks) and so give no contribution. Thus, the
results (\ref{2loop.anom})--(\ref{d}) can be directly used for
flavor-conserving magnetic moment operators as well.\footnote{We would
like to thank Matthias Jamin for bringing this fact to our attention.}

\section{The \bsg case}

	Having in mind the potential application of our results to the
calculation of the next-to-leading logarithmic corrections to the \Bsg
decay rate, we now rewrite the $2 \times 2$ anomalous dimension matrix
of the magnetic moment operators once again in the conventional
notation used in the analyses of this process. If we normalize the two
magnetic moment operators as in ref.~\cite{bh}
\bea \label{ope}
Q_7  & = & \f{e}{8 \pi^2} m_b \bar{s}_{\alpha} \sigma^{\mu \nu}
(1+\gamma_5) b^{\alpha} F_{\mu \nu}            \vspace{0.2cm} \\
Q_8 & = & \f{g}{8  \pi^2} m_b \bar{s}_{\alpha} \sigma^{\mu \nu}
(1+\gamma_5) (T^a)^{\alpha}_{\beta} b^{\beta} G_{\mu \nu}^a
\eea
then their $2 \times 2$ anomalous dimension matrix can be obtained
from the corresponding entries in eqs.~(\ref{2loop.anom})--(\ref{a})
just by subtracting $\gamma_m$ from the diagonal terms. At the
two-loop level, the anomalous dimension of the mass reads
\cite{Pascual-Tarrach} with our sign convention
\be
\gamma_m = - \f{g^2}{16 \pi^2} 6 C_f + \f{g^4}{(16 \pi^2)^2} C_f
\left( - \f{97}{3} N - 3 C_f +\f{10}{3} f \right) + {\cal O}(g^6).
\ee
Consequently, the anomalous dimension matrix for $Q_7$ and $Q_8$ is
\be \label{gamma}
\hat{\gamma}(g) = \hat{\gamma}^{(0)} \f{g^2}{16 \pi^2} +
		  \hat{\gamma}^{(1)} \f{g^4}{(16 \pi^2)^2}
+ {\cal O}(g^6),
\ee
where
\be \label{gamma0}
\hat{\gamma}^{(0)} =
\left( \begin{array}{lc}
8 C_f & 0\\
8 C_f Q_s & 16 C_f - 4 N \\
\end{array} \right) =
\left( \begin{array}{rc}
\f{32}{3} &  0        \\
-\f{32}{9}  & \f{28}{3} \\
\end{array} \right)
\ee
and, in the \MS scheme,
\begin{eqnarray} \label{gamma1}
\hat{\gamma}^{(1)} =
\left( \begin{array}{cc}
\phantom{Q_s} C_f \left(\f{548}{9} N - 16 C_f - \f{56}{9} f \right) &
0 \\
Q_s C_f \left(\f{404}{9} N - 32 C_f - \f{56}{9} f \right) &
-\f{458}{9} -\f{12}{N^2}+ \f{214}{9} N^2 + \f{56}{9} \f{f}{N} -
\f{13}{9} f N \\
\end{array} \right) = \nonumber
\end{eqnarray}
\be
=\left( \begin{array}{rc}
\f{4688}{27} &  0        \\
-\f{2192}{81}  & \f{4063}{27} \\
\end{array} \right).
\ee

	Unfortunately, knowing the above anomalous dimension matrix is
not enough to write the full next-to-leading RGE for the coefficients
of $Q_7$ and $Q_8$. As explained in detail in ref.~\cite{bh}, the
three-loop mixing of the four-quark operators into the magnetic moment
operators gives contributions of the same order as the two-loop
self-mixing of the latter operators. Therefore, our calculation is an
important step towards the next-to-leading calculation of the \Bsg
rate, but by no means completes this calculation. We hope that the
methods described in the present paper can provide helpful tools in
calculating the desired three-loop mixing.

	The numerical value of the next-to-leading corrections we have
calculated can be easily determined from the matrix in
eq.~(\ref{gamma1}) and the analysis presented in ref.~\cite{bh}. The
equations (4)--(9) of that paper summarize the leading-order
expressions for the \bsg decay rate which is proportional to the
square of the quantity denoted there by $C_7^{(0)eff}$. Inserting the
matrices from eqs.~(\ref{gamma0}) and (\ref{gamma1}) above into the
equations (21)--(28) of ref.~\cite{bh}, we find the next-to-leading
contribution that adds to $C_7^{(0)eff}$ in the \MS-scheme:
\bea
\label{dc7}
\lefteqn{\Delta C_7^{eff}(\mu) = \f{\al(M_W)}{4 \pi} \left[ \f{37208}{4761}
\left( \eta^{\f{16}{23}} - \eta^{-\f{7}{23}} \right) C_7^{(0)}(M_W) +
\right.} \nonumber \\
\nonumber \\
& + & \left. \left( \f{297664}{14283} \eta^{-\f{7}{23}}
-\f{7164416}{357075} \eta^{-\f{9}{23}} +\f{256868}{14283}
\eta^{\f{14}{23}} -\f{6698884}{357075} \eta^{\f{16}{23}} \right)
C_8^{(0)}(M_W) \right],
\eea
with $\eta$, $C_7^{(0)}(M_W)$ and $C_8^{(0)}(M_W)$ defined as in
eqs.~(5), (8) and (9) of ref.~\cite{bh}. For $\al(M_Z) = 0.12$, $m_t =
174$ GeV and $\mu = 5$ GeV, the above contribution causes a decrease
of the predicted branching ratio by around 5\%. However, no
phenomenological conclusions can be drawn before all the other
next-to-leading corrections are calculated. Only then the total
correction is renormalization-scheme independent.

	The leading order expression for the coefficient of the
gluonic operator is given in the last equation of ref.~\cite{bh}. The
next-to-leading contribution from the two-loop self-mixing of $Q_8$
reads (in the \MS-scheme)
\be \label{dc8}
\Delta C_8^{eff}(\mu) = \f{\al(M_W)}{4 \pi} \f{64217}{9522}
\left( \eta^{\f{14}{23}} - \eta^{-\f{9}{23}} \right) C_8^{(0)}(M_W)
\ee
which in the Standard Model case gives numerically around 2.4\%
decrease of the coefficient, i.e.~a similar change as in the photonic
case. Even when the full next-to-leading calculation is completed, its
effect on the coefficient of the gluonic operator is not expected to
have any measurable physical consequences. So we should treat
eq.~(\ref{dc8}) only as a by-product of the calculation performed for
the photonic operator $Q_7$.

	In the end, we would like to point out once again that our
two-loop anomalous dimension matrix is insensitive to the scheme used
for $\gamma_5$ in dimensional regularization. Two-loop mixing usually
depends on the scheme used for $\gamma_5$, and the scheme dependence
of the two-loop anomalous dimension matrix cancels in physical
quantities with the scheme dependence of the one-loop matrix elements
\cite{66,ql,Pisa,bsgmix}. In the case of $Q_7$--$Q_8$ mixing, we have
completely eliminated $\gamma_5$ from the calculation by using
chirality conservation in QCD and QED interactions. This was
equivalent to a trivial application of the recent Pisa scheme for
$\gamma_5$ \cite{Pisa}. However, exactly the same anomalous dimension
matrix would be obtained in any of the other commonly used schemes HV,
NDR and DRED (see Buras and Weisz in ref. \cite{66} for a summary of
these schemes). So the situation here is much simpler than in the case
of four-quark operators.

	To conclude: We have calculated the two-loop QCD mixing of the
magnetic moment operators which are the only EOM-nonvanishing
dimension-five flavor-changing operators that arise in the Standard
Model after integrating out the $W$, $Z$ and Higgs bosons and the
top-quark. Our calculation is an important contribution to the
phenomenologically desired next-to-leading QCD corrections to the
\bsg decay. This was the last two-loop QCD mixing to be calculated for
the effective hamiltonian describing flavor changing processes at low
energies in the Standard Model and many of its extensions.

\setcounter{secnumdepth}{0}
\section{Acknowledgements}

	We would like to thank Andrzej Buras, Matthias Jamin, Manfred
Lindner, Uli Nierste, Hubert Simma and Peter Weisz for helpful
discussions. This research has been supported by the German
Bundesministerium f\"ur Forschung und Technologie under contract 06 TM
732, by the CEC Science project SC1-CT91-0729 and by the Polish
Committee for Scientific Research.

\section{Appendix}

	In order to derive the equations (\ref{111})--(\ref{n11}) one
starts from the integral
\be
J_{211}(M,m_1,m_2) = \int \f{d^D q}{(2 \pi)^D} \f{d^D r}{(2 \pi)^D}
          \f{1}{[q^2 - M^2]^2 [r^2 - m_1^2] [(q+r)^2 - m_2^2]}.
\ee
Differentiating the integrand with respect to $m_1$ or $m_2$ we find
an integral that is convergent by Weinberg's theorem \cite{Weinberg}
(i.e.~all the subdiagrams of the corresponding scalar diagram have
negative degree of divergence). Consequently, the divergent part of
$J_{211}(M,m_1,m_2)$ is independent of $m_1$ and $m_2$. We can
therefore replace $J_{211}(M,m_1,m_2)$ by $J_{211}(M,0,0)$  and easily
calculate the latter with help of the Feynman parametrization
\bea
J_{211}(M,m_1,m_2) & = & J_{211}(M,0,0) + {\rm convergent\;\;terms} =
\nonumber\\
 & = & -\f{1}{(4\pi)^4} \left[ \f{2}{\epsilon^2} + \f{1}{\epsilon} \left( 1
- 2 \gamma_E - 2 \ln\f{M^2}{4 \pi} \right) \right] + {\rm
convergent\;\;terms}
\eea
By integrating and/or differentiating the above result with respect to
the masses appropriately many times and then setting all the masses
equal, we obtain the equations (\ref{111})--(\ref{n11}). This is also
how one learns that $I_{abc}$ is convergent when two of its indices
are greater than 1 and all of them are positive.

	The short derivation presented in this appendix is motivated
by the appendix of ref.~\cite{Veltman} where also the finite parts of
the integrals are given. However, we disagree with the global sign of
the explicit expression for $(M,M | M_1 | M_2)$ given there.

\setlength{\baselineskip}{0.2in}
\setlength{\textheight}{9.4in}

\end{document}